# Action principle and weak invariants

Sumiyoshi Abe [a-d, *] and Congjie Ou [a]

[a] *Physics Division, College of Information Science and Engineering,
   Huaqiao University, Xiamen 361021, China*
[b] *ESIEA, 9 Rue Vesale, Paris 75005, France*
[c] *Department of Physical Engineering, Mie University, Mie 514-8507, Japan*
[d] *Institute of Physics, Kazan Federal University, Kazan 420008, Russia*

**Abstract**   A weak invariant associated with a master equation is characterized in such a way that its spectrum is not constant in time but its expectation value is conserved under time evolution generated by the master equation. Here, an intriguing relationship between the concept of weak invariants and the action principle for master equations based on the auxiliary operator formalism is revealed. It is shown that the auxiliary operator can be thought of as a weak invariant.

---

* Corresponding author.

Invariants play fundamental roles in physics. In particular, the profound connection between continuous symmetries and conservation laws is widely known as Noether's theorem [1] and is indispensable for developing the action integral for dynamics. However, a situation becomes somewhat involved if a system is not autonomous. In this case, as long as the dynamics is unitary and is generated by a Hamiltonian with explicit time dependence, still there exists an invariant, which depends on time but whose spectrum is constant. Such an invariant is referred to as a *strong* invariant. A celebrated example is the Lewis-Riesenfeld invariant of the time-dependent harmonic oscillator [2] (see also Ref. [3]). It is known that Lewis-Riesenfeld strong invariant can be obtained through quantization of the classical invariant derived from the action principle and Noether's theorem [4]. Strong invariants of the various kinds have been applied to diverse problems including geometric phases [5-7], the coherent and squeezed states [8-10], quantum computation [11], nonstationary quantum field theory in time-dependent backgrounds [12-14] and quantum cosmology [15].

Recently, the concept of *weak* invariants has been developed for nonunitary subdynamics [16]. In particular, the Lewis-Riesenfeld strong invariant of the time-dependent quantum harmonic oscillator has been generalized to the case of the time-dependent quantum damped harmonic oscillator, and the weak invariant (which is different from the quantity studied in Ref. [17]) has explicitly been constructed within the framework of the master equation of the Lindblad type [18,19]. In contrast to a strong invariant, the spectrum of a weak invariant depends on time, but its expectation value is conserved. An origin of its physical usefulness is concerned with a new concept in finite-time quantum thermodynamics, that is, the isoenergetic process interacting with the energy bath, along which the internal energy is conserved [20,21]. There, the relevant weak invariant is the time-dependent Hamiltonian of a subsystem under consideration. It is also worth emphasizing that such a process should be distinguished from the isothermal process familiar in classical thermodynamics because of the quantum-mechanical violation of the law of equipartition of energy.

In this article, we reveal a hidden connection between the action principle for a general master equation without time reversal invariance and a weak invariant.

Consider a master equation

$$i\frac{\partial \hat{\rho}}{\partial t} = \pounds(\hat{\rho}) \tag{1}$$

as an initial value problem, where $\hat{\rho} = \hat{\rho}(t)$ is a density operator describing an open quantum system, and $\pounds$ is a certain linear superoperator that may depend explicitly on time, in general. Here and hereafter, $\hbar$ is set equal to unity for the sake of simplicity. A weak invariant, $\hat{I} = \hat{I}(t)$, associated with Eq. (1) is defined as a solution of the following equation:

$$i\frac{\partial \hat{I}}{\partial t} + \pounds^*(\hat{I}) = 0, \tag{2}$$

where $\pounds^*$ stands for the adjoint of $\pounds$. Then, it is straightforward to see that the expectation value, $\langle \hat{I} \rangle = \mathrm{tr}(\hat{I}\hat{\rho})$, is constant in time: $d\langle \hat{I} \rangle / dt = 0$.

In the special case of the Lindblad equation [18,19], the superoperators in Eqs. (1)

and (2) read

$$\pounds(\hat{\rho}) = [\hat{H}, \hat{\rho}] - i \sum_n \alpha_n \left( \hat{L}_n^\dagger \hat{L}_n \hat{\rho} + \hat{\rho} \hat{L}_n^\dagger \hat{L}_n - 2 \hat{L}_n \hat{\rho} \hat{L}_n^\dagger \right), \quad (3)$$

$$\pounds^*(\hat{I}) = -[\hat{H}, \hat{I}] - i \sum_n \alpha_n \left( \hat{L}_n^\dagger \hat{L}_n \hat{I} + \hat{I} \hat{L}_n^\dagger \hat{L}_n - 2 \hat{L}_n^\dagger \hat{I} \hat{L}_n \right), \quad (4)$$

respectively, where the subsystem Hamiltonian $\hat{H}$, the nonnegative $c$-numbers $\alpha_n$'s, and the Lindbladian operators $\hat{L}_n$'s may also depend explicitly on time. The first term on the right-hand side in Eq. (3) is the unitary part that appears in the Liouville-von Neumann equation, whereas the second term is called the dissipator responsible for nonunitarity of the dynamics. From Eq. (4), it can be shown [16] that the spectrum of $\hat{I}$ is, in fact, time-dependent.

Here, a comment is made on the adjoint superoperator in Eq. (2). In general, it is desirable that $\pounds^*$ has the following property:

$$\pounds^*(\hat{A} + c) = \pounds^*(\hat{A}), \quad (5)$$

where $\hat{A}$ is a certain operator and $c$ is any $c$-number. In fact, Eq. (4) satisfies this condition. The case of time-independent $c$ is special since a weak invariant shifted by such a constant is clearly a weak invariant, and Eq. (2) remains unchanged under the constant shift if Eq. (5) is satisfied.

Now, the purpose of the present work is to elucidate how the weak invariant satisfying Eq. (2) is connected to the action principle for the master equation in Eq. (1).

As in the case of the Lindblad equation, the master equation describing the subdynamics does not possess time reversal invariance. Therefore, to construct the action integral for the equation, it is convenient to extend the space of variables. Thus, following the work in Ref. [22], we introduce an auxiliary operator $\hat{\Lambda} = \hat{\Lambda}(t)$ and consider the following action integral:

$$S[\hat{\rho}, \hat{\Lambda}] = -\int_{t_i}^{t_f} dt \left\langle \dot{\hat{\Lambda}} - i\, \pounds^*\left(\hat{\Lambda}\right)\right\rangle - \left\langle \hat{\Lambda} \right\rangle\bigg|_{t=t_i}, \qquad (6)$$

where $t_i$ ($t_f$) is the initial (final) time and the overdot stands for differentiation with respect to time. Unlike the density operator, the auxiliary operator has to be neither positive semi-definite nor normalized, in general.

The following remark is made about the action integral given above. To calculate its variation with respect to $\hat{\rho}$, the normalization condition on it should be taken into account. However, it turns out not to be necessary to add such a constraint to the action integral. The reason is as follows. Let us redefine the auxiliary operator as follows:

$$\hat{\Lambda}(t) \to \hat{\Lambda}(t) + \int_t^{t_f} ds\, \lambda(s), \qquad (7)$$

where $\lambda$ is a $c$-number function. Then, the action integral in Eq. (6) is rewritten as

$$S[\hat{\rho}, \hat{\Lambda}] \to S[\hat{\rho}, \hat{\Lambda}] + \int_{t_i}^{t_f} dt\, \lambda(t) \left( \operatorname{tr} \hat{\rho}(t) - \operatorname{tr} \hat{\rho}(t_i) \right). \tag{8}$$

In this form, $\lambda$ is seen to play a role of the Lagrange multiplier associated with the normalization condition: that is, $\operatorname{tr} \hat{\rho}(t) = 1$ holds if the initial density operator is fixed to be normalized $\operatorname{tr} \hat{\rho}(t_i) = 1$. Therefore, it is actually not necessary to add the constraint on the normalization condition to the action integral in Eq. (6). Furthermore, it should be noted that the "transformation" in Eq. (7) keeps the final condition $\hat{\Lambda}(t_f)$ unchanged.

Now, variations with respect to $\hat{\rho}$ and $\hat{\Lambda}$ yield

$$\delta_\rho S[\hat{\rho}, \hat{\Lambda}] = -\int_{t_i}^{t_f} dt\, \operatorname{tr}\left\{ \left[ \dot{\hat{\Lambda}} - i\, \pounds^*(\hat{\Lambda}) \right] \delta \hat{\rho} \right\} - \operatorname{tr}\left( \hat{\Lambda}(t_i)\, \delta \hat{\rho}(t_i) \right), \tag{9}$$

$$\delta_\Lambda S[\hat{\rho}, \hat{\Lambda}] = \int_{t_i}^{t_f} dt\, \operatorname{tr}\left\{ \left[ \dot{\hat{\rho}} + i\, \pounds(\hat{\rho}) \right] \delta \hat{\Lambda} \right\} - \operatorname{tr}\left( \hat{\rho}(t_f)\, \delta \hat{\Lambda}(t_f) \right), \tag{10}$$

respectively. Thus, under the fixed initial and final conditions, $\delta \hat{\rho}(t_i) = 0$ and $\delta \hat{\Lambda}(t_f) = 0$, we obtain, from Eqs. (9) and (10), that

$$i\frac{\partial \hat{\Lambda}}{\partial t} + \pounds^*(\hat{\Lambda}) = 0, \tag{11}$$

as well as the master equation in Eq. (1). The pairwise structure of the fixed initial and final conditions respectively on the density and auxiliary operators is characteristic of the action principle based on the auxiliary operator formalism [22]. It may be interpreted as the restoration of time reversal invariance in the extended space of variables (if the superoperators do not have explicit time dependence).

Consequently, comparing Eq. (11) with Eq. (2), we find that an auxiliary operator in the action principle for a master equation is a weak invariant.

At the very beginning of this article, we have mentioned the relation between Noether's theorem and (strong) invariants. The corresponding problem for weak invariants will be discussed elsewhere.

**Acknowledgements**

The authors would like to thank the support by a grant from National Natural Science Foundation of China (No. 11775084). The work of S.A. has also been supported in part by a Grant-in-Aid for Scientific Research from the Japan Society for the Promotion of Science (No. 16K05484), the program of Fujian Province, China, and the program of Competitive Growth of Kazan Federal University from the Ministry of Education and Science, Russian Federation. This work has been completed while S.A. has stayed at the Wigner Research Centre for Physics with the support of the Distinguished Guest Fellowship of the Hungarian Academy of Sciences. He would like to thank the Wigner Research Centre for Physics for the warm hospitality extended to him.